\def\gtorder{\mathrel{\raise.3ex\hbox{$>$}\mkern-14mu
		\lower0.6ex\hbox{$\sim$}}}
   \def\ltorder{\mathrel{\raise.3ex\hbox{$<$}\mkern-14mu
		\lower0.6ex\hbox{$\sim$}}}

   \def\artanh{\mathrel{\rm arctanh}}

   \documentstyle[12pt,aaspp4]{article}
    \begin{document}

   \title{A Survey for Large Image-Separation Lensed Quasars}

   \author{\bf Dan Maoz$^{1}$,
       Hans-Walter Rix$^{2,3}$, Avishay Gal-Yam$^{1}$, and Andrew Gould$^{4}$}

   $^1$ School of Physics \& Astronomy and Wise Observatory,

    Tel-Aviv University, Tel-Aviv 69978, Israel.
      
         dani@wise.tau.ac.il

   $^2$ Steward Observatory, University of Arizona, Tucson, AZ 85721

        rix@as.arizona.edu

$^3$ Alfred P. Sloan Fellow

$^4$ Department of Astronomy, Ohio State University, Columbus, OH 43210

        gould@payne.mps.ohio-state.edu

   \begin{abstract}
The statistics of gravitationally lensed quasars with multiple
images in the $0.1''-7''$ 
range have been measured in various surveys.
 Little is known, however, about lensed-quasar
 statistics at larger image separations, which probe masses
on the scale of galaxy clusters. We extend the results
of the {\it Hubble Space Telescope} ({\it HST})
Snapshot Survey for Lensed Quasars 
to the $7''-50''$ range for a sub-sample of 76 quasars that is free
of known selection effects.
Using a combination of multicolor photometry and spectroscopy, we show
that none of the point sources in the entire field of view of the {\it HST} 
observations of these quasars are lensed images. Large-separation quasar lensing
is therefore not common. We carry out a detailed calculation of the
expected statistics of large-separation lensing for this quasar sample,
 incorporating 
realistic input for the mass profiles and mass function of galaxy clusters.
We find that the observational null results are consistent with the expected effect of
galaxy clusters, even if these have existed in their present form and number
since $z\sim 2$ (and certainly if they were formed more recently). The rarity
of large-separation lensed quasars can rule out some extreme scenarios, e.g. that
the mass-function of clusters has been severely underestimated, or that 
large mass concentrations that are not associated with galaxies (i.e. ``failed''
clusters) are common. The rareness of cluster lensing also sets limits
on the cosmological constant $\lambda$ that are independent of limits derived from galaxy
lensing. The lensing frequency depends strongly on the central density
of clusters. The lensing statistics
of larger quasar samples (e.g. the Sloan Digital Sky Survey)
can probe the structure, number,
and evolution of clusters, as well as the geometry of space.
   \end{abstract}


\centerline{submitted to {\it The Astrophysical Journal}: December 15, 1996}

   \section{Introduction}
The statistics of gravitational lensing can provide a powerful probe
of the geometry and the mass content of the universe
out to large redshifts (e.g. Refsdal 1964; Press \& Gunn 1973). 
Turner, Ostriker, \& Gott (1984) first explored
lensing probabilities due to galaxies, and the resulting
image separation distributions.
The {\it Hubble Space Telescope}
({\it HST}) Snapshot Survey for lensed quasars (Bahcall et al. 1992; Maoz et al. 1992; 1993a;
1993b) was the first such large survey of a well-defined sample of 498 quasars.
Exploiting the angular resolution of {\it HST}, it showed that about 1\%
of luminous quasars at $z>1$ are gravitationally lensed into multiple
images with separations in the $0.1''-7''$ range. Maoz \& Rix (1993) used 
the Snapshot Survey results to demonstrate that early-type galaxies must
have, on average, dark massive halos similar to those of spiral galaxies,
and that the geometry of the Universe is not dominated by a cosmological
constant $\lambda$, setting an upper limit of $\lambda < 0.7$. 
Ground-based surveys of 360 additional quasars and their analysis
 (see Kochanek 1996, and references
therein) have confirmed these results.
While the statistics of gravitationally lensed quasars with multiple
images in the angular range expected due to galaxy lensing have been
probed by  the Snapshot and other surveys, little is known about lensed-quasar
 statistics at larger image separations, which probe masses
on the scale of galaxy clusters. There are no confirmed cases of quasar
splitting with separations above $7''$.\footnote{ Saunders et al. (1997)
recently suggested that PC1643+4631A\&B ($198''$ separation)
are images of a single quasar lensed by a massive $z\sim1$ cluster,
despite the fact that there is a small redshift difference in the spectra.
Our results in the present work suggest that lensing with such image
separations in highly unlikely.}

Narayan \& White (1988), Cen et al. (1994), Wambsganss et al. (1995),
Kochanek (1995), and Flores \& Primack (1996) have all compared large-scale structure
formation models to the observed statistics of large-separation lensed quasars.
However, the statistics utilized were the known lensed quasars in
published catalogs, which are basically literature compilations.
As emphasized by Kochanek (1995), most quasars found in quasar 
surveys are near the faint detection limits of the surveys. The
surveys will therefore
generally not find faint lensed images of a given quasar, unless
the two images are close in brightness. Kochanek (1995) estimates
that that the completeness level of the catalogs, in terms
of large-separation lenses, is only 20\%. Furthermore, Gould, Bahcall,
\& Maoz (1993) have shown that quasar surveys using spectroscopic
selection methods are biased against quasars having stars
nearby in projection. Presumably, the same bias operates against neighboring
lensed images, and would further select against inclusion of
lensed quasars in the catalogs. There is therefore a possibility that
large-separation lensing is more common than assumed, and that
many such lensed quasars have been missed.

In the first part of this paper, we present the results of the first extensive survey for
large-separation lensed quasars among known quasars. We use multi-color
photometry and spectroscopy to test whether each of the point sources in the entire
$70''\times 70''$ 
field of view of the {\it HST} Planetary Camera (PC) exposures of 76 quasars 
in the original Snapshot Survey could be lensed images of the quasars.
In the second part of the paper, we carry out a calculation of the
expected lensing statistics for this particular sample and its observational
parameters. The calculation follows closely that of Maoz \& Rix (1993)
and Rix et al. (1995)
for small-separation lensing by galaxies, with galaxy clusters playing the
previous role of galaxies. In addition to including effects such
as magnification bias and observational detection limits, our calculation
uses a realistic cluster mass profile that is motivated by N-body
simulations (Navarro, Frenk, \& White 1995, 1996, 1997) and is
consistent with the observations of 
``radial arcs'' in clusters (Bartelmann 1996). The choice of mass profile
is important, since lensing calculations are sensitive to
the presence of a core vs. a singular profile (e.g. Flores \& Primack 1996).

\section{Sample and Observations}
In the {\it HST} Snapshot Survey for lensed quasars, 498 quasars were
imaged with the Planetary Camera. The sample consisted of most of the quasars
in the V\'eron-Cetty \& V\'eron (1989) catalog with redshift $z>1$,
absolute magnitude $M_V< -25.5$ ($H_0=100$ km s$^{-1}$
Mpc$^{-1}$, $q_0=0.5$), and galactic latitude $|b|>10^{\circ}$.
(See Maoz et al. 1993b, for further details.) Gould, Bahcall, \& Maoz (1993)
catalogued all the point sources apearing in the PC exposure of 
each quasar to a typical limiting magnitude of $V\sim 21.3$ mag
(see Gould, Bahcall, \& Flynn 1996),
 and used them to study Galactic structure. Gould et al. (1993) found
that quasars discovered by spectroscopic means (e.g. objective prisms) tended
to ``avoid'' foreground stars out to separations of $40''$.
The reason for this is unclear.
 The same
selection effect may operate on lensed images of the quasar,
such that spectroscopically selected quasars are less likely to
be lensed at separations above several arcseconds. Gould et al. (1993)
defined an ``unbiased'' sample of 166 quasar fields in which the quasar
was discovered by non-spectroscopic means (color-excess, radio, X-rays).
For the present study, we have chosen from among the unbiased sample
the 88 quasars in the anti-center Galactic hemisphere (i.e. $90^{\circ}<l<270^{\circ}$,
where $l$ is Galactic longitude). By looking only at quasars in the 
anti-center hemisphere, we greatly reduce the number of stars that
have to be checked to see if they are lensed images.

There are no point sources in the fields of 29 of the anti-center quasars,
so these
automatically pass the test for not being lensed (within the detection
limits and the field of view probed by a given exposure).
In nine additional quasar fields, the only other sources present are significantly
brighter than the quasar. If they were lensed images of the quasars,
they would have been identified as such by the original surveys.
(This is not necessarily true of X-ray surveys, which may have
poor angular resolution. However, the only objects around an X-ray selected
quasar that were rejected based on this criterion are two $V\sim 12$ mag stars
near the $V\sim 18$ mag quasar 0438$-$166.)
We carried out $V$ and $I$ CCD photometry of the quasars and the point sources fainter
than them in the remaining 50 quasar fields. Useful measurements
were obtained for 38 of the fields.
 This leaves us with an observed sample of 76 quasars.\footnote{A slight
bias is introduced here, since isolated quasars are automatically 
included the sample, whereas some quasars surrounded with point sources
are excluded. The bias could be corrected by eliminating from the
sample a corresponding number (seven) of the isolated quasars. This
would reduce the sample and the lensing predictions for it by 10\%,
 not affecting any of our conclusions.}
The 76 quasars and their parameters are listed in Table 1.

 The $V$ and $I$ observations were made
on 1993 March 4--9 at the Kitt Peak National Observatory (KPNO)
and on 1993 September 5--10 at the Cerro Tololo Inter-American 
Observatory (CTIO). Both runs used Tektronix $1024\time1024$-pixel CCDs
at the Cassegrain focus of 0.9 m telescopes. Landolt (1992) standards
were observed throughout the nights. Conditions were photometric,
with a scatter of less than 0.03 mag around the fits to the Landolt
magnitudes. The ground-based $V$ and $I$ magnitudes of each point source
appearing in the
{\it HST} exposures were measured using the Daophot point-source-function (PSF)
fitting routine (Stetson 1987) within IRAF\footnote{IRAF (Image Reduction and Analysis
Facility) is distributed by the National Optical Astronomy Observatories,
which are operated by Aura, Inc., under cooperative agreement with the
National Science Foundation.}. Errors were calculated by
combining in quadrature the error in the photometric solution,
as determined from its covariance matrix, the scatter around the
photometric error, and the Daophot PSF-fitting error.

 The $V$ magnitude and $V-I$ color
of each quasar and faint point source in the fields of the un-isolated
 quasars is listed in Table 2. The first line for each field gives the
results for the quasar in the field, and the subsequent lines for the
stars. Positions of the stars are given in Gould et al. (1993).
The $V-I$ uncertainty listed in Table 2 does not include the absolute
photometric calibration error, since this error cancels out
in the {\it difference} between the $V-I$ color of a quasar and a star observed
in the same CCD frame.

  From a comparison of $V-I$ colors
between each quasar and its neighboring objects, we can reject all stars
in the fields of 26 quasars as candidate lensed images.
The star near one quasar, 0024+22, can be
rejected as a lensed image based on the absence of a radio counterpart (Condon et al. 1981),
as described in Maoz et al. (1993a).
The remaining 14 stars around 11 quasars
have a color difference between star and quasar of $\Delta (V-I) < 0.3$ mag. Since
such color differences among lensed images are possible due to differential
reddening in the different light paths, these cases were kept for further
testing. We measured $B$ and $V$ magnitudes for these remaining candidates on
1994 November 11 and 12 at the Wise Observatory 1m telescope
 with a Tektronix $1024\times 1024 $
back-illuminated CCD. The reduction and calibration was as for the KPNO
and CTIO observations described above. The $B-V$ colors are given in Table 2. 

If a $V-I$ color difference between lensed images is due to differential reddening,
then the expected $B-V$ color difference among the images will be $\Delta (B-V) \approx
0.625 \Delta (V-I)$ (Rieke \& Lebofsky 1985).
Based on $\Delta (B-V)$ significantly greater (after accounting for all the
 measurement errors) than expected from $\Delta (V-I)$ and reddening, we
 excluded eight of these stars,
leaving six point sources in the fields of six quasars, each having
both $V-I$ and $B-V$ colors similar to the quasar in their field.
Spectra were obtained for these 6 sources at the Multiple-Mirror
Telescope (MMT) on 1995, November 25--27, using a 1200 l/mm grating,
covering 4500 \AA\ to 6000 \AA\ at 2 \AA\ resolution. The spectra show that all six
are foreground stars. They are marked as such in the right-hand 
column in Table 2.

We have thus demonstrated that none of the point sources detected
in the PC field of view of 76 unbiased Snapshot Survey quasars in the anti-center
hemisphere are lensed quasar images.

\section{Calculation of Lensing by Clusters}

\subsection{Algorithm}
We have carried out a calculation of the expected number and 
distribution in angular separation of lensed quasars due to
the effects of intervening galaxy clusters. Our calculation
follows closely that of Maoz \& Rix (1993). Briefly, for a 
given observed quasar and a particular lens (i.e.,
 a cylindrically-symmetric 
cluster of a particular
redshift and mass) we find the ``critical radius'' inside
which lensing into multiple images occurs. For every impact
parameter inside this radius, we calculate the three image positions
and their magnifications.
We weight
the  image distribution for the particular
lens according to the magnification bias, the cross section 
for lensing at that redshift, the number density of clusters
of that mass, and the volume of space included in a redshift 
interval. We then integrate numerically the image distributions
over grids in cluster mass and redshift (from $z=0$ to $z$ of the quasar),
 to obtain the probability
that a given quasar in the sample is lensed, as a function of image
separation.  We weight this distribution
for each quasar according to the detection efficiency as a function
of image separation.
These probability distributions are calculated
for every quasar in the sample and added, to give the 
expected number, and distribution in image separation, of lensed 
quasars in the survey. We provide more details below,
with emphasis on places where the calculation differs from
Maoz \& Rix (1993).

\subsection{Cluster Mass Profile}

To model the mass profile of galaxy clusters, we have used 
the radial mass density function
\begin{equation}
\rho(x)=\frac{\rho_s}{x(1+x)^2},
\end{equation}
where the radial coordinate, $x$, is in units of the scale
radius $r_s$, $x\equiv r/r_s$. Navarro et al. (1995, 1996, 1997) 
have found that this mass profile describes well the dark-matter
halos produced in cosmological N-body
simulations for a variety of initial density fluctuation
spectra. Note that at its center, the profile is singular,
with $\rho\propto r^{-1}$. It is thus intermediate between the
two types of mass profiles that have been considered in previous
works on cluster lensing statitics, the singular isothermal sphere
models, where $\rho\propto r^{-2}$, and models with a core,
where $\rho$ flattens to a constant at the cluster center.
It is similar at small radii to the Hernquist profile,
$\rho\propto r^{-1}(r+a)^{-3}$, considered by Flores \& Primack (1996),
but falls off more gently at large radii.
Bartelmann (1996) has shown that the mass profile of equation 1
can produce radially distorted images of background sources,
so-called ``radial arcs'', whose existence were previously
thought to indicate cores in galaxy clusters. Flores \& Primack (1996)
have shown that, if clusters have large cores, the number of
large-angle splittings is greatly reduced, even in cosmological
models with excessive large-scale structure (e.g. standard Cold
Dark Matter [CDM]). They have argued that large-separation lensing is
therefore not a sensitive probe of large-scale structure. 
Bartelmann's demonstration that clusters with a central density
singularity can produce radial arcs renews that possibility that 
such clusters are efficient splitters of background quasars,
and hence useful probes of large-scale structure and its evolution.
In view of the results of Navarro et al. and Bartelmann, we consider
equation 1 to be the one of the more realistic cluster mass profiles.

For our lensing calculations, we have used the expressions given
by Bartelmann (1996) for the mass $m(x)$ enclosed within a cylinder
of radius $x$. The bending angle of a light ray passing the
cluster at impact parameter $x$, and hence the image positions
and magnifications, are determined by $m(x)/x$ and its derivative
(see, e.g., Maoz \& Rix 1993). The mathematical details
for the present case are provided in an Appendix.

From Figure 9 of Navarro et al. (1997) we estimate that
that the scale length $r_s$ depends on the cluster mass $M$
as 
\begin{equation}
r_s=300\left(\frac{M}{10^{15}M_\odot}\right)^{\gamma} h^{-1} {\rm kpc},
\end{equation}
 where $h$ is the Hubble constant in units
of 100 km s$^{-1}$ Mpc$^{-1}$.
The index $\gamma$ varies among cosmological models between
$\gamma \sim 1/3$ (CDM) to $\gamma \sim 1$ (for a model with initial density
fluctuation power spectrum of power-law form with index $n=0$).
When $\gamma$ is large, low-mass clusters have dense central regions
and can become efficient lenses.

\subsection{Observational Detection Limits}

As in Maoz \& Rix (1993), we have incorporated the detection
limits for multiple images for each quasar into the calculation.
In Maoz \& Rix (1993), the detection limits included the
angular resolution limit for detection of images with a given
brightness ratio, and the flux limit of the exposure. In the 
present work, we have also included the incompleteness of 
the survey at large angular separations due to the positioning
of the quasar on the PC field of view. For technical reasons,
the quasar was generally 
not in the center of the PC field of view. The angular range
from $0''-70''$ was therefore covered with
a varying degree of completeness for each angle and each quasar.
For every quasar,
the lensing probability 
distribution  at angle $\theta$ was scaled by a factor equal
to the fraction of a circle, of radius $2\pi\theta$
with center at the quasar position, that
is within the PC field of view. The position of each
quasar on the PC is given in Table 1.
See Maoz et al. (1993) for more details on the PC observations.
The effect of this angular selection function on
the completeness of the survey is examined in \S 4.

\subsection{Magnification Bias}
The magnification bias is the over-representation
of lensed quasars in a flux-limited quasar sample
due to the facts that lensing increases the apparent
brightness of a quasar and that there are more faint quasars
than bright ones. The magnification bias
is calculated as in Maoz \& Rix (1993), with
the probability $P(A)$ for amplification by a 
factor $A$ calculated for the cluster lenses under
consideration here. We have updated the slopes
of the quasar luminosity function to $\alpha=-1.3\pm0.2$,
 $\beta=-3.87\pm0.15$, the absolute magnitude of the
break in the local luminosity function to $M_{Q0}^*=-20.87\pm0.25$
and the luminosity evolution power-law index to
$k_L=3.2\pm0.1$,
according to the results of Boyle et al. (1990).

\subsection{Cluster Mass Function and Evolution}
To represent the number density of clusters
of a given mass, we have used  the observational
 results of Bahcall \& Cen (1993), rather than 
the theoretical large-scale structure predictions
used by previous studies of large-separation lensing.
Bahcall \& Cen find that, for groups and clusters of galaxies with
mass $M$ between $10^{13}$ and $10^{15}M_{\odot}$, the 
mass function can be represented analytically by
$n(>M)=4\times 10^{-5} (M/M^*)^{-1}{\rm exp}(-M/M^*)h^3$ Mpc$^{-3}$.
Here $M^*=(1.8\pm0.3)\times 10^{14}h^{-1} M_{\odot}$,and $M$ is the mass
within a $1.5h^{-1}$ Mpc radius sphere around the 
cluster center.

 Bahcall \& Cen also show that the
extrapolation of this function to $10^{12}M_{\odot}$ has
a value similar to the space density of ``noncluster'' $L^*$ {\it galaxies},
and note that this continuity is expected from general 
theoretical expectations (e.g. Press \& Schechter 1974).
Since the observed galaxy luminosity
function actually has an exponential cutoff at $L*$, the implication would
be that
dark halos form at all masses, but when the mass is above $10^{12}M_{\odot}$, a group
or a cluster will form in it, rather than a galaxy. In practice,
it is unknown whether or not dark halos in the $10^{12}-10^{13}M_{\odot}$ range
exist. Since, as will turn out, halos in this mass range can be important
for large separation lensing, we will allow for two possibilities:
that the cluster mass function can be extrapolated down to $10^{12}M_{\odot}$, as in
Bahcall \& Cen, or that there is a lower cutoff to the mass of clusters
at $10^{13}M_{\odot}$ (i.e. a gap in the halo mass function, contrary
to Press-Schechter theory). We will further examine the consequences of
such a gap in \S 4.

 The derivative of $n(>M)$
with respect to $M$ gives $n_0(M)$, the local ($z=0$) space density of clusters
of mass $M$ to $M+dM$. As a first approximation,
we have assumed that the number of clusters per co-moving volume element
does not evolve
between $z=0$ and the redshift of the quasar (typically $z\sim 2$),
i.e. the number density at $z$ is $n(z)=n_0 (1+z)^3$.
Flores \& Primack (1996) show that the no-evolution assumption
is a reasonable approximation to the numerical CDM calculation
of Cen et al. (1994), and occurs because a cluster-sized
perturbation is already virialized inside a radius of
 $200(v/1000~{\rm km~s}^{-1})h^{-1}$ kpc by $z=3$. It is the inner
parts of the cluster which determine the lensing statistics.
In non-CDM models, clusters may form more
recently, and one would expect that at higher $z$ there
are fewer clusters and/or the mass in clusters is less concentrated.
However, we find that even with the simplistic assumption of
no evolution, the expected number of large-separation lensed quasars
in the survey is generally $<1$, and hence consistent with the observed 
null result. Clearly, incorporation of recent cluster formation
would further lower the prediction. More details
on the results and their dependence on the input parameters
are given below.

\section{Results and Discussion}
We have calculated the number of lensed quasars we expect
to detect in our sample
as a function of image separation for a variety of combinations
of the input parameters of clusters. Figure 1 shows the 
expected distribution for a particular choice of parameters.
Also shown is the distribution that would result if the angular 
selection function resulting from the positioning of the quasar 
on the PC did not exist. We see that the angular selection
function does not seriously impair our ability to detect lensed
quasars, and has some effect only at the largest ($>30''$)
separations. The total number of expected lensed quasars above $7''$ separation
in this particular model is 0.032 (0.041 without the angular selection
function), consistent with our null result.

 Although the clusters produce
lensed quasars with separations smaller than $7''$ too, comparison to
the observations in the $0''-7''$
 range is complicated because lensing with such
separations is produced by galaxy lenses as well. A case in point is
the lensed quasar 0957+561, with $6.1''$ image separation.
It was not observed with {\it HST} because
it was previosly known to be lensed but, in principle,
it is part of the Snapshot sample (see Maoz et al. 1992). Since it is
radio-discovered and in the anti-center hemisphere, it would be included
in the present sample as well. However, a cluster and a galaxy both play
a significant role in lensing this quasar. It would therefore be unclear
whether or not it counts as one detection when comparing to the predictions
of lensing by clusters. By comparing models and observations only for
 separations $>7''$, we restrict ourselves to lensing by clusters alone.

 A free parameter
in our calculation is $\gamma$, the power-law index relating
the scale radius $r_s$ to the total cluster mass (eq. 2). Navarro et al.
(1997) predict that $\gamma$ varies between cosmological models
in the range of 1/3 to 1. Among the input parameters, the total
number of lensed quasars is, by far, most sensitive to $\gamma$.
A large $\gamma$  gives low- and intermediate-mass clusters
a large central density, turning them into efficient lenses.
Figure 2 shows the lensing distribution on a logarithmic
scale for various values of $\gamma$. The models with
$\gamma \ltorder 0.7$ produce $<<1$ lensed quasars in our
sample. These models are therefore consistent with our
observed null result, even if the other input parameters,
such as the number density of clusters or the mass of  
an $M^*$ cluster have both been  underestimated by an
order of magnitude.

On the other hand, models with $\gamma \sim 1$ produce
about one expected lens for the ``standard'' input parameters
(the Bahcall \& Cen 1993 mass function and an Einstein-de Sitter $\Omega=1$
cosmology). Models predicting 3 (4.5) or more lenses can be
rejected at $>95\%$ ($>99\%$) confidence based on Poisson
statistics. Factors of a few in the predicted number of 
lenses can result from mild changes in the parameters of the
mass function (which is empirically not well constrained)
or the cosmology (e.g., lowering $\Omega$, or introducing
a cosmological constant $\lambda$ will raise the
prediction). Some models with high $\gamma$ and various
combinations of the other parameters can therefore be
rejected. 

Figure 3 shows the dependence of the lensing distribution
on $M^*$, the exponential upper mass cutoff
 in the cluster mass function.
For low $\gamma$, the total number and the mean image separation
of lensed quasars both increase with $M^*$. At higher $\gamma$ (not
shown),
the total number of lensed quasars is insensitive to $M^*$, and
only the centroid of the distribution shifts slowly with $M^*$.
This occurs because, for large $\gamma$, the more massive 
clusters that are introduced by raising $M^*$ have large scale
lengths, and so do not lens effectively.

The parameter plane of $\gamma$ and $M^*$ is shown in
Figure 4, which gives the total number of predicted
lensed quasars with $>7''$ separation
for combinations of these parameters. $\Omega=1$ and
the Bahcall \& Cen (1993) value of $n(M^*)$,
 the number density of $M^*$-mass clusters,
are assumed.
Increasing $n(M^*)$ would increase the number of lensed
quasars by the same proportion. We
see that, if $\gamma$ is large, the number density
of clusters cannot be much above the Bahcall \& Cen (1993)
estimate, or some lensed quasars would have been found
in the survey. A great advantage of gravitational 
lensing is that it probes mass, rather than light. The above
result therefore also shows that, unless $\gamma$ is small,
there cannot be a large population of ``failed clusters'',
i.e. dark halos not containing clusters of galaxies.
The predicted number of lensed quasars approximately
doubles when going from a flat $\Omega=1$ to an open $\Omega=0$
model. Low-density open models with $\gamma \sim 1$ predict
more than 3 lenses, so are inconsistent with the data.

Because dense, low-mass clusters can be efficient lenses,
introducing a lower mass cutoff at $10^{13}M_{\odot}$
in the cluster mass function (see \S 3.5) greatly reduces
the lensing prediction. For example, in a $\gamma=0.8$
model, the expected lensing distribution
decreases by a factor of $\sim 30$ below $10''$, and by a factor
of $\sim 4$ above $10''$, so the total expected number of lensed quasars
decreases
by about a factor of 15. For $\gamma=0.67$ (i.e., low-mass
clusters are not so dense) the distribution
 decreases by a factor of $\sim 2$
below $10''$, but is unchanged above $10''$.

Finally, high-$\gamma$ models in a flat Universe dominated
by a cosmological constant can also be ruled out. This
is shown in Figure 5, displaying the $\lambda - \gamma$
plane. For example, a model with $\gamma=1$ and $\lambda=0.7$
produces over 5 lensed quasars in our sample. This limit
on $\lambda$ is independent of the limits that have
been derived based on small-separation quasar lensing
by galaxies (Fukugita \& Turner 1990; Maoz \& Rix 1993;
Kochanek 1996). Fukugita \& Peebles (1993) and
Malhotra, Rhoads, \& Turner (1997) have suggested that
small-separation lensing statistics
can be reconciled with a $\lambda$-dominated Universe by invoking
dust in the lensing galaxies. The excess number of lensed
quasars would then be hidden by extinction. This argument
is not applicable to lensing by clusters. Maoz (1995) has shown
that rich clusters do not significantly redden quasars
that are behind them. A similar demonstration has been made
for poorer clusters by Williams \& Hawkins (1996). 
Conversely, if the dependence of cluster scale length
on cluster mass is weak (i.e. if $\gamma$ is small), or if
there is a significant decrease in the number of clusters
between now and $z\sim2$, or if there is a gap in the
halo mass function between masses of $\sim 10^{12}-10^{13}M_{\odot}$,
a $\lambda$-dominated Universe is allowed by the present
large-separation lensing statistics.

Due to the smallness of this first large-separation sample,
we have limited the range of models we have examined, and
avoided complications such as evolution in cluster number
and structure. Upcoming surveys, such as the Sloan Digital
Sky Survey, will detect and test orders of magnitude more quasars for
lensing by galaxies and by clusters. If  $\gamma$ is
small, we predict that even these surveys will find no
examples of large-separation lensed quasars. If, on the
other hand, clusters have dense centers, many such lenses
will be found. Their detailed statistics can then serve
as a valuable probe of the structure, number,
and evolution of galaxy clusters, and of the geometry of the Universe.

\acknowledgements
We thank J. Wambsganss
for suggesting this project. 

\appendix

\section{The Lensing Equation for Clusters}

A light ray passing with
an impact parameter $b$ from the center of a cylindrically symmetric
mass distribution is bent by an angle (e.g. Weinberg 1972)
\begin{equation}
\alpha=\frac{4GM(<b)}{c^2b} ,
\end{equation}
where $M(<b)$ is the total mass that is projected inside of $b$.
The lensing equation, relating the angle $\theta_I$ between the lens
 and the projected image to the angle $\theta_S$ between the source
 and the lens, is
\begin{equation}
\theta_{S}=\frac{D_{LS}}{D_{OS}}\alpha(\theta_I)-\theta_I ,
\end{equation}
where $D_{LS}$ and $D_{OS}$ are the angular diameter distances between
the lens and the source and between the observer and the source,
respectively. The condition for gravitationally lensing  a source 
into multiple images
is that the source be projected on the sky within an angle
$\theta<\theta_{cr}$ of the lens,
\begin{equation}
\theta_{cr}=\frac{D_{LS}}{D_{OS}}\alpha(\theta_1)-\theta_1 .
\end{equation}
where $\theta_1$ is defined by
\begin{equation}
\frac{D_{LS}}{D_{OS}}\frac{d \alpha}{d\theta}\bigg |_{\theta=\theta_1}=1 .
\end{equation}

For a lensing cluster with 
surface density profile $\Sigma(r)$ and scale radius $r_s$,
Equation (A1) can be rewritten as:
\begin{equation}
\alpha=\frac{4GM_{1.5}}{c^2 r_s} f(x),
\end{equation}
where $M_{1.5}$ is the mass enclosed within $1.5 h^{-1}$ Mpc, 
the dimensionless function $f(x)$ is defined as 
\begin{equation}
f(x)\equiv \frac{1}{x}\frac{\int_0^x \Sigma(r)r dr}{\int_0^{1.5{\rm Mpc}} \Sigma(r)r dr}
=\frac{1}{x}\frac{g(x)}{g(1.5 {\rm Mpc}/r_s)},
\end{equation}
and 
\begin{equation}
x\equiv \frac{b}{r_s}.
\end{equation}
$g(x)$ for the mass profile of equation 1 is given by Bartelmann (1996) 
as
\begin{equation}
  g(x) = \ln{x\over2} + \cases{
  {2\over\sqrt{x^2-1}}\arctan\sqrt{x-1\over x+1} & $(x>1)$ \cr
  {2\over\sqrt{1-x^2}}\artanh\sqrt{1-x\over 1+x} & $(x<1)$ \cr
  1 & $(x=1)$ \cr
  }\;.
\end{equation}
Combining equations (A3) and (A5), we can write the critical radius
for lensing as:
\begin{equation}
r_{cr}=r_s \left( \frac{\Sigma_{av}}{\Sigma_{cr}}f(x_1)-x_1 \right).
\end{equation} 
Here $\Sigma_{av}\equiv M_{1.5}/\pi r_s^2$,
 $\Sigma_{cr}$ is the critical surface mass density, defined as
\begin{equation}
\Sigma_{cr}\equiv \frac{c^2}{4\pi G}\frac{D_{OS}}{D_{OL}D_{LS}},
\end{equation}
where $D_{OL}$ is the angular diameter distance from the observer to
the lens, and $x_1=D_{OL}\theta_1/r_s$, so that, from equation (A4),
\begin{equation}
\frac{df}{dx}\bigg |_{x_1}=\frac{\Sigma_{cr}}{\Sigma_{av}}.
\end{equation}

We show in Figure A1
the functions  $g(x)/x$ and  $d(g(x)/x)/dx$.
Determining the critical radius for
a cluster of given mass involves solving numerically 
equation (A11) for $x_1$ and substituting $x_1$ into equation (A9).



\begin{deluxetable}{lcccccccc}
\scriptsize
\tablewidth{370pt}
\tablecaption{Unbiased Anticenter Snapshot Quasar Sample}
\tablehead{\colhead{Quasar}&\colhead{Other Name$^1$}&\colhead{Sel$^2$}&\colhead{$l^{II}$}&
\colhead {$b^{II}$}&\colhead{$V^3$}&\colhead{$z$}& \colhead{$-M_V^4$}&\colhead{PC position$^5$} }
\startdata
0004$+$171   &        &   R& 108.03&$-$44.11& 18.7& 2.89& 25.9& 6(550,595) \nl
0024$+$22    &        &   C& 115.66&$-$39.83& 17.0& 1.11& 25.6& 8(174,629) \nl
0033$+$0951  & 4C09.01&   R& 116.84&$-$52.56& 17.7& 1.92& 26.0& 6(304,308) \nl
0058$+$0155  & PHL938 &   C& 127.75&$-$60.59& 17.1& 1.93& 26.6& 7(190,552) \nl
0100$-$270   &        &   R& 205.78&$-$87.41& 18.1& 1.60& 25.2& 8(186,149) \nl
0109$+$17    &        &   R& 129.76&$-$44.70& 18.7& 2.15& 25.3& 7(184,211) \nl
0119$-$04    &        &   R& 142.30&$-$66.06& 17.0& 1.95& 26.8& 6(285,310) \nl
0122$-$00    &        &   R& 141.16&$-$61.76& 16.6& 1.07& 25.9& 8(628,073) \nl
0132$+$20    &        &   C& 136.45&$-$40.96& 17.9& 1.78& 25.7& 6(343,318) \nl
0136$+$176   &        &   R& 194.15&$-$78.45& 18.7& 2.73& 25.7& 6(389,352) \nl
0136$-$231   &        &   R& 138.85&$-$43.49& 17.8& 1.89& 25.9& 6(121,278) \nl
0151$+$0448  & PHL1222&   C& 150.40&$-$54.46& 17.9& 1.91& 25.8& 6(152,219) \nl
0215$+$165   &        &   R& 151.18&$-$41.27& 17.5& 1.90& 26.2& 6(251,344) \nl
0220$-$142   &        &   R& 185.79&$-$65.00& 18.8& 2.43& 25.4& 6(239,223) \nl
0225$-$014   &        &   R& 168.86&$-$55.26& 18.6& 2.03& 25.2& 7(266,300) \nl
0226$-$038   &        &   R& 171.90&$-$56.93& 17.2& 2.06& 26.7& 6(242,259) \nl
0229$+$13    &        &   R& 157.09&$-$42.74& 17.9& 2.07& 26.0& 6(258,258) \nl
0232$-$04    &        &   R& 174.46&$-$56.16& 16.3& 1.43& 26.8& 6(246,225) \nl
0244$-$128   &        &   R& 190.42&$-$59.32& 18.4& 2.2 & 25.6& 6(232,260) \nl
0256$-$005   &        &   R& 177.19&$-$49.23& 17.5& 1.99& 26.3& 8(362,061) \nl
0302$-$223   &        &   X& 211.08&$-$59.39& 16.9& 1.41& 26.2& 6(167,150) \nl
0329$-$255   &        &   R& 219.43&$-$54.30& 17.8& 2.69& 26.6& 6(250,214) \nl
0335$-$336   &        &   X& 233.40&$-$53.87& 18.5& 2.27& 25.6& 6(169,258) \nl
0347$-$241   &        &   R& 218.73&$-$49.98& 17.1& 1.88& 26.6& 8(734,115) \nl
0355$-$48    &        &   R& 256.16&$-$48.45& 16.6& 1.01& 25.8& 6(240,261) \nl
0402$-$362   &        &   R& 237.74&$-$48.48& 16.9& 1.42& 26.2& 6(220,241) \nl
0438$-$166   &        &   X& 214.14&$-$36.25& 17.9& 1.96& 25.9& 6(253,237) \nl
0438$-$43    &        &   R& 248.41&$-$41.57& 19.5& 2.85& 25.0& 6(248,258) \nl
0448$-$392   &        &   R& 242.77&$-$39.61& 16.7& 1.29& 26.2& 6(227,237) \nl
0451$-$28    &        &   R& 229.02&$-$37.02& 18.2& 2.56& 26.1& 6(243,262) \nl
0636$+$68    &        &   R& 147.17&   24.15& 16.6& 3.17& 28.2& 6(105,133) \nl
0642$-$349   &        &   R& 244.29&$-$16.49& 18.0& 2.17& 26.0& 7(262,160) \nl
0731$+$65    &        &   R& 151.00&   29.12& 18.2& 3.03& 26.5& 6(246,245) \nl
0759$+$341   &        &   R& 187.12&   28.74& 18.5& 2.44& 25.7& 6(066,344) \nl
0804$+$4959  & OJ508  &   R& 169.16&   32.56& 18.3& 1.43& 24.8& 6(422,265) \nl
0808$+$28    &        &   R& 193.47&   29.13& 18.3& 1.91& 25.4& 6(252,264) \nl
0812$+$33A   &        &   R& 188.89&   31.10& 19.2& 2.42& 25.0& 6(271,226) \nl
0827$+$24    &        &   R& 200.02&   31.88& 17.5& 2.06& 26.4& 6(256,226) \nl
0830$+$115   &        &   R& 213.99&   27.71& 18.0& 2.97& 26.6& 6(236,260) \nl
0831$+$101   &        &   R& 215.61&   27.42& 19.6& 1.76& 23.9& 8(378,109) \nl
0836$+$1932  & 4C19.31&   R& 206.13&   32.10& 17.6& 1.69& 25.9& 8(633,180) \nl
0838$+$4536  & US1498 &   C& 174.94&   37.97& 17.5& 1.41& 25.6& 6(052,202) \nl
0843$+$1339  & 4C13.39&   R& 213.25&   31.36& 17.5& 1.88& 26.2& 6(047,276) \nl
0848$+$1533  & LB8755 &   C& 211.77&   33.24& 17.8& 2.01& 26.0& 6(265,237) \nl
0848$+$1623  & LB8775 &   C& 210.93&   33.74& 17.6& 1.93& 26.1& 6(250,227) \nl
0854$+$1907  & LB8956 &   C& 208.45&   36.00& 18.0& 1.89& 25.7& 6(240,251) \nl
0856$+$124   &        &   R& 216.18&   33.93& 19.4& 1.76& 24.1& 8(293,096) \nl
0903$+$175   &        &   X& 211.22&   37.49& 18.0& 2.77& 26.5& 6(255,198) \nl
0907$+$381   &        &   R& 185.04&   42.97& 17.6& 2.16& 26.4& 6(248,271) \nl
0932$+$367   &        &   R& 187.39&   47.80& 18.4& 2.84& 26.1& 6(281,176) \nl
0945$+$114   &        &   R& 224.00&   44.12& 18.9& 2.14& 25.0& 6(079,300) \nl
0945$+$4337  & US987  &   C& 176.79&   49.90& 18.2& 1.89& 25.5& 6(138,087) \nl
0946$+$301   &        &   C& 197.83&   50.24& 16.2& 1.22& 26.6& 6(297,288) \nl
0955$+$4739  & OK492  &   R& 170.06&   50.73& 18.7& 1.87& 25.0& 6(255,252) \nl
1008$+$133   &        &   C& 225.40&   50.05& 16.3& 1.29& 26.6& 6(217,353) \nl
1011$+$091   &        &   X& 231.51&   48.54& 17.7& 2.26& 26.4& 6(249,223) \nl
1038$+$065   & 4C06.41&   R& 241.11&   52.65& 16.6& 1.27& 26.3& 6(243,264) \nl
1038$+$528   &        &   R& 157.50&   54.97& 18.6& 2.30& 25.5& 7(615,291) \nl
1045$+$60    & 4C60.15&   R& 146.72&   50.93& 18.7& 1.72& 24.8& 8(074,330) \nl
1137$+$30    & US2778 &   C& 197.29&   74.12& 16.7& 1.57& 26.6& 7(265,653) \nl
1139$+$2833  & US2828 &   C& 204.60&   74.70& 17.3& 1.61& 26.1& 6(470,667) \nl
1148$+$38    &        &   R& 167.15&   73.11& 17.2& 1.30& 25.7& 6(232,251) \nl
1206$+$459   &        &   C& 144.63&   69.62& 15.5& 1.15& 27.1& 6(237,228) \nl
1211$+$33    &        &   R& 173.89&   79.92& 17.6& 1.60& 25.7& 6(247,243) \nl
1215$+$33    &        &   R& 171.72&   80.56& 18.1& 2.60& 26.2& 6(180,177) \nl
1215$+$6423  & 4C64.15&   R& 128.99&   52.61& 18.1& 1.29& 24.8& 7(232,122) \nl
1246$+$3746  & BSO1   &   C& 125.76&   79.61& 17.2& 1.24& 25.6& 5(652,231) \nl
1248$+$401   &        &   C& 123.48&   77.27& 16.1& 1.03& 26.3& 6(513,361) \nl
1257$+$34    & B201   &   C& 109.54&   82.52& 16.7& 1.37& 26.3& 6(237,254) \nl
1259$+$3427  & BSO6   &   C& 105.93&   82.60& 18.0& 1.95& 25.8& 8(283,078) \nl
1309$+$3402  & BSO8   &   C&  91.00&   82.06& 17.7& 1.75& 25.8& 7(289,092) \nl
1323$+$6530  & 4C65.15&   R& 117.22&   51.50& 17.5& 1.62& 25.9& 6(393,134) \nl
1356$+$5806  & 4C58.29&   R& 106.58&   57.09& 17.2& 1.37& 25.8& 6(266,344) \nl
2251$+$24    &        &   R&  91.71&$-$30.91& 18.5& 2.33& 25.6& 7(345,288) \nl
2345$+$061   &        &   R&  96.24&$-$53.16& 18.1& 1.54& 25.2& 6(251,256) \nl
2353$+$154   &        &   R& 103.82&$-$45.11& 18.3& 1.80& 25.3& 6(039,303) \nl
\enddata
\tablecomments{
\nl
$^1$\ Listed for objects for which
V\'eron-Cetty and V\'eron (1989) give a name that does not include
the coordinates.\nl
$^2$\ Quasar discovery method: R-radio; C-color;
X-x-ray.\nl
$^3$\ $V$ magnitude determined from the
HST exposure, accurate to $\approx \pm
 0.1$ magnitudes.\nl
$^4$\ Absolute $V$ magnitude,
 assuming $H_0=100$ km s$^{-1}$
Mpc$^{-1}$, $q_0=0.5$.\nl
$^5$\ Planetary Camera (PC) CCD number (5
through 8) and pixel coordinate with the
origin at the PC apex.
}
\end{deluxetable}

\begin{deluxetable}{cccccccc}
\scriptsize
\tablewidth{300pt}
\tablecaption{Quasar Fields with Stars}
\tablehead{\colhead{QSO Field}&\colhead{$V$}&\colhead{$\sigma$}&\colhead{$V-I$}&
\colhead {$\sigma$}&\colhead{$B-V$}&\colhead {$\sigma$}& \colhead {Comments} }
\startdata
0004$+$17 &  18.607&  0.019&  0.611&  0.022&      &      &          \nl
          &  20.396&  0.049&  1.549&  0.061&      &      &          \nl
          &  19.755&  0.029&  1.692&  0.033&      &      &          \nl
0024$+$22 &  16.583&  0.015&  0.509&  0.014&      &      &          \nl
          &  19.208&  0.023&  1.038&  0.027&      &      &          \nl
          &  19.971&  0.043&  0.758&  0.060&      &      &   1      \nl
0119$-$04 &  17.199&  0.015&  0.993&  0.014& 0.450& 0.015&          \nl
          &  20.645&  0.050&  0.695&  0.078& 0.156& 0.324&   2      \nl
0122$-$00 &  16.747&  0.022&  0.305&  0.037&      &      &          \nl
          &  19.771&  0.037&  1.369&  0.057&      &      &          \nl
0220$-$14 &  18.776&  0.024&  0.603&  0.030&      &      &          \nl
          &  20.314&  0.050&  0.980&  0.064&      &      &          \nl
0225$-$01 &  18.485&  0.017&  0.715&  0.020& 0.408& 0.038&          \nl
          &  20.033&  0.037&  0.853&  0.053& 0.615& 0.145&   2      \nl
0226$-$03 &  17.444&  0.017&  0.642&  0.018&      &      &          \nl
          &  20.951&  0.077&  1.389&  0.095&      &      &          \nl
0229$+$13 &  18.253&  0.019&  1.005&  0.021& 0.197& 0.026&          \nl
          &  20.401&  0.049&  1.285&  0.062& 1.490& 0.489&          \nl
0232$-$04 &  16.404&  0.014&  0.660&  0.012&      &      &          \nl
          &  20.866&  0.062&  1.851&  0.070&      &      &          \nl
0302$-$22 &  16.641&  0.014&  0.692&  0.012& 0.284& 0.008&          \nl
          &  18.943&  0.022&  1.785&  0.024&      &      &          \nl
          &  20.327&  0.042&  0.722&  0.062& 0.481& 0.191&   2      \nl
0329$-$25 &  17.742&  0.017&  0.422&  0.018&      &      &          \nl
          &  20.238&  0.042&  2.345&  0.043&      &      &          \nl
0335$-$33 &  18.415&  0.020&  1.046&  0.024&      &      &          \nl
          &  19.260&  0.031&  1.547&  0.034&      &      &          \nl
0347$-$24 &  17.148&  0.016&  0.673&  0.025&      &      &          \nl
          &  21.221&  0.057&  1.997&  0.067&      &      &          \nl
0438$-$16 &  17.907&  0.016&  0.660&  0.010&$-.075$& 0.012&          \nl
          &  20.116&  0.036&  0.602&  0.052& 0.279& 0.099&          \nl
0448$-$39 &  16.818&  0.019&  0.680&  0.021&      &      &          \nl
          &  17.443&  0.021&  2.714&  0.022&      &      &          \nl
0451$-$28 &  17.944&  0.016&  0.481&  0.018&      &      &          \nl
          &  19.921&  0.036&  0.862&  0.052&      &      &          \nl
0636$+$68 &  16.464&  0.021&  0.675&  0.025& 0.848& 0.016&          \nl
          &  17.147&  0.022&  0.675&  0.027& 0.542& 0.023&          \nl
          &  19.086&  0.048&  0.715&  0.058& 0.682& 0.105&  2       \nl
          &  18.522&  0.033&  2.091&  0.036&      &      &          \nl
0642$-$34 &  17.857&  0.015&  0.841&  0.010& 0.368& 0.023&          \nl
          &  18.581&  0.017&  0.894&  0.014& 0.822& 0.042&          \nl
          &  19.681&  0.025&  1.832&  0.024&      &      &          \nl
          &  20.600&  0.046&  0.836&  0.067& 0.370& 0.176&  2       \nl
0731$+$65 &  18.125&  0.017&  0.432&  0.029&      &      &          \nl
          &  19.894&  0.040&  1.199&  0.064&      &      &          \nl
0804$+$49 &  18.205&  0.040&  0.693&  0.049&      &      &          \nl
          &  20.538&  0.114&  2.464&  0.116&      &      &          \nl
0808$+$28 &  18.094&  0.025&  0.727&  0.027& 0.234& 0.009&          \nl
          &  20.102&  0.073&  0.952&  0.085& 0.795& 0.081&          \nl
          &  18.599&  0.029&  1.147&  0.030&      &      &          \nl
0827$+$24 &  17.494&  0.036&  0.356&  0.039&      &      &          \nl
          &  18.098&  0.041&  2.131&  0.043&      &      &          \nl
          &  20.331&  0.160&  1.927&  0.163&      &      &          \nl
0830$+$11 &  18.010&  0.080&  0.380&  0.082&      &      &          \nl
          &  20.350&  0.080&  1.623&  0.088&      &      &          \nl
0831$+$10 &  19.261&  0.068&  0.729&  0.091&      &      &          \nl
          &  19.272&  0.059&  2.073&  0.067&      &      &          \nl
0848$+$16 &  17.542&  0.040&  0.655&  0.051&      &      &          \nl
          &  19.265&  0.111&  2.325&  0.115&      &      &          \nl
0903$+$17 &  17.985&  0.045&  1.141&  0.055&      &      &          \nl
          &  19.803&  0.112&  2.278&  0.118&      &      &          \nl
0932$+$36 &  18.703&  0.031&  0.614&  0.037&      &      &          \nl
          &  20.498&  0.120&  1.218&  0.135&      &      &          \nl
1008$+$13 &  16.370&  0.027&  0.561&  0.038&      &      &          \nl
          &  17.116&  0.030&  1.475&  0.039&      &      &          \nl
1011$+$09 &  17.649&  0.043&  1.285&  0.051&      &      &          \nl
          &  19.471&  0.057&  0.040&  0.116&      &      &          \nl
1038$+$06 &  16.449&  0.027&  0.392&  0.036&      &      &          \nl
          &  17.370&  0.032&  1.961&  0.039&      &      &          \nl
1148$+$38 &  17.125&  0.014&  0.636&  0.005& 0.325& 0.007&          \nl
          &  20.341&  0.070&  0.760&  0.081& 0.527& 0.101&  2       \nl
          &  20.993&  0.115&  2.588&  0.116&      &      &          \nl
1211$+$33 &  17.451&  0.026&  0.657&  0.034& 0.152& 0.013&          \nl
          &  18.780&  0.057&  0.678&  0.079& 0.649& 0.033&  3       \nl
          &  19.739&  0.125&  0.717&  0.170& 0.649& 0.033&  3       \nl
1248$+$40 &  16.235&  0.021&  0.430&  0.026&      &      &          \nl
          &  20.507&  0.102&  1.708&  0.109&      &      &          \nl
1309$+$34 &  17.581&  0.029&  0.530&  0.029&      &      &          \nl
          &  21.190&  0.110&  1.209&  0.136&      &      &          \nl
          &  19.850&  0.064&  2.697&  0.064&      &      &          \nl
1323$+$65 &  17.356&  0.025&  0.603&  0.036&      &      &          \nl
          &  18.479&  0.038&  1.114&  0.049&      &      &          \nl
2251$+$24 &  18.176&  0.019&  1.006&  0.021& 0.722& 0.032&          \nl
          &  18.588&  0.021&  1.512&  0.022&      &      &          \nl
          &  19.375&  0.028&  1.311&  0.036& 1.581& 0.142&          \nl
2345$+$06 &  18.417&  0.018&  0.812&  0.022&      &      &          \nl
          &  19.777&  0.029&  1.970&  0.036&      &      &          \nl
2353$+$15 &  18.034&  0.018&  0.801&  0.029&      &      &          \nl
          &  19.730&  0.033&  1.425&  0.046&      &      &          \nl
\enddata
\tablecomments{First line for every field gives results for quasar, following
lines are for stars in the field fainter than the quasar. Positions of the
stars are given in Gould et al. (1993).\nl
$1$-\ Not a lensed image, based on absence of radio emission,
see text and Maoz et al. (1993a).\nl
$2$-\ Star, based on spectroscopy with the MMT.\nl
$3$-\ $1''$-separation binary, see Gould et al. (1995). $B-V$ color is for the
combined pair.
}
\end{deluxetable}

\begin{figure}
\epsscale{1.}
\plotone{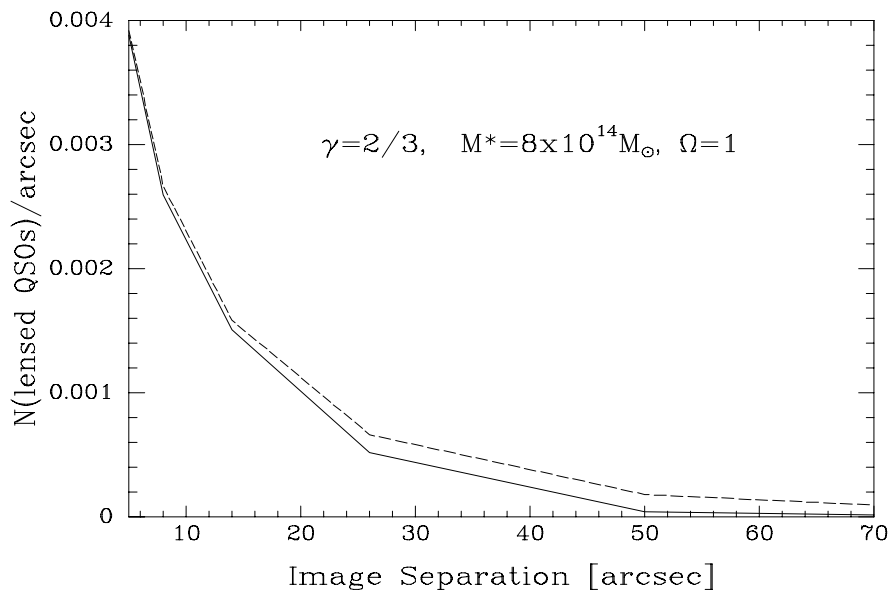}
\caption{ The 
expected distribution of lensed quasars for a particular choice of parameters.
$\gamma$ is the power law index relating cluster scale length
and mass. $M^*$ is the exponential 
cutoff mass in the cluster mass function, here assumed to have
a value 4 times the Bahcall \& Cen (1993) estimate.
Dashed line is the distribution that would result if the angular 
selection function resulting from the positioning of the quasar 
on the detector did not exist. The total number of expected lensed quasars
with separation $>7''$ in this particular model is 0.032 (0.041
 without the angular selection
function), consistent with our null result. }
\end{figure}

\begin{figure}
\epsscale{1.}
\plotone{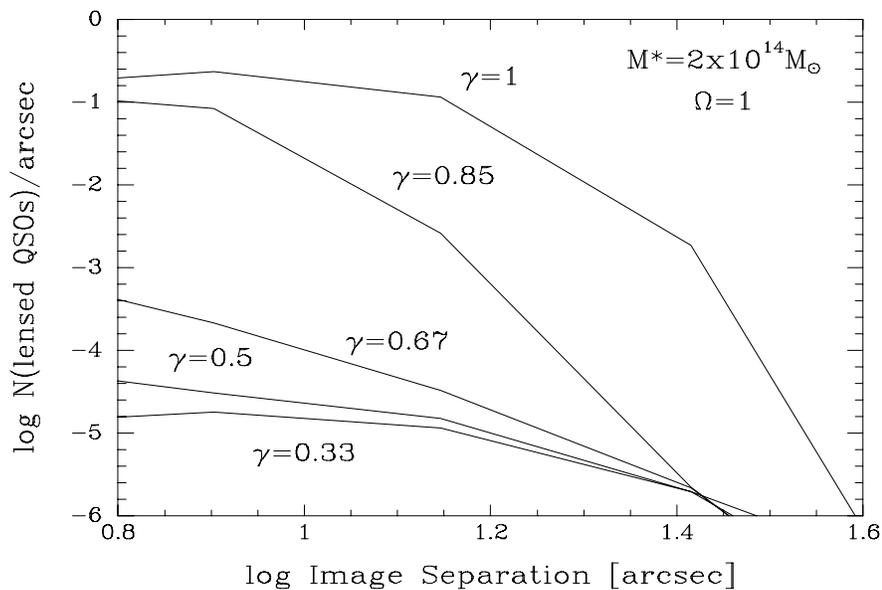}
\caption{The expected number of lensed quasars
 for various values of $\gamma$,
 shown on a logarithmic
scale. The models with
$\gamma \ltorder 0.7$ produce $<<1$ lensed quasars in our
sample, and are therefore consistent with our
observed null result, even if the other input parameters,
such as the number density of clusters or the mass of a 
an $M^*$ cluster have both been underestimated by an
order of magnitude.
}
\end{figure}

\begin{figure}
\epsscale{1.}
\plotone{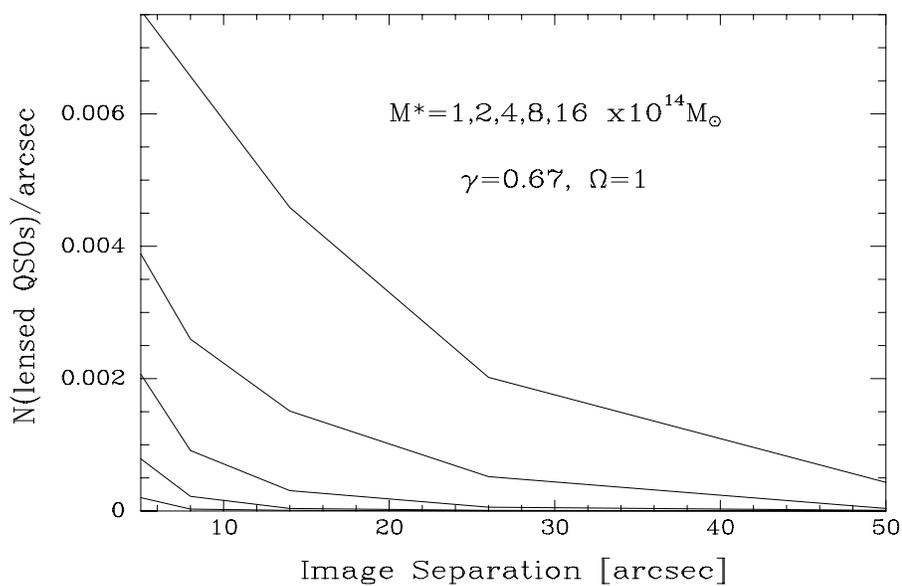}
\caption{The dependence of the lensing distribution
on $M^*$, the cutoff mass  in the cluster mass function.
The higher curves have higher $M^*$; the second curve
from the bottom corresponds to the Bahcall \& Cen (1993) value.
For low $\gamma$, such as assumed in this example ($\gamma=2/3$),
 the total number and the mean image separation
of lensed quasars both increase with $M^*$.
}
\end{figure}

\begin{figure}
\epsscale{1.}
\plotone{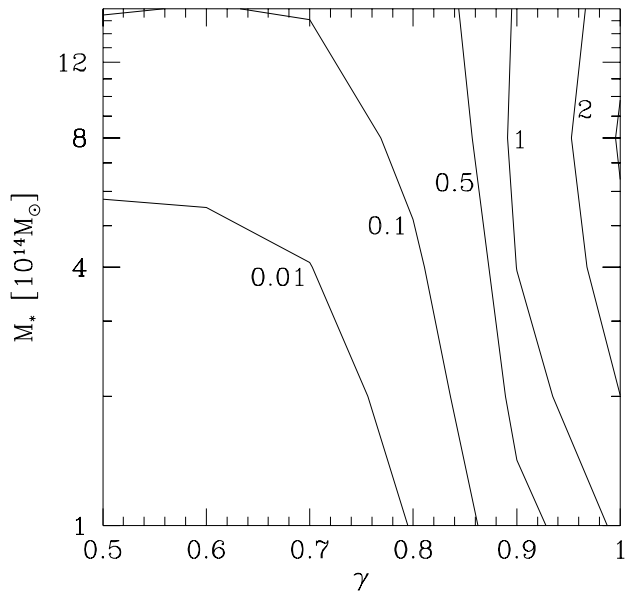}
\caption{Total number of lensed quasars with $>7''$ separation
 expected for combinations
of $\gamma$, the index of the power-law relating cluster scale length
to mass, and $M^*$, the cutoff mass in the cluster mass function.
 The Bahcall \& Cen
(1993) value for $M^*$ is $2\times 10^{14}h^{-1} M_{\odot}$.
 $\Omega=1$ is assumed; lowering $\Omega$
to 0 approximately doubles the predicted number of lensed quasars.
Models predicting $\ge 3$ ($\ge 4.5$) lenses can be rejected at 95\% (99\%)
confidence.
}
\end{figure}

\begin{figure}
\epsscale{1.}
\plotone{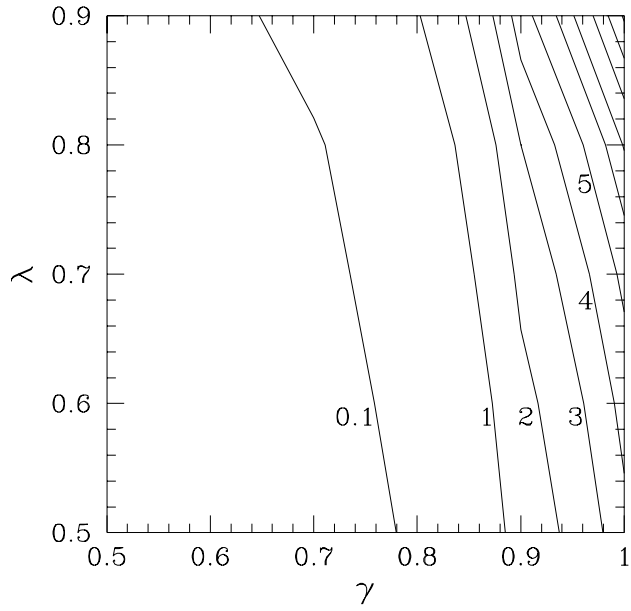}
\caption{Same as fig. 4, but for combinations
of $\gamma$ and
$\lambda$, the cosmological constant in dimensionless form,
for flat ($\Omega + \lambda=1$) cosmologies.
High-$\lambda$ models in which low-mass clusters have
large central mass concentration (i.e., high $\gamma$) produce
several large-separation lenses, and can be rejected.
}
\end{figure}

\begin{figure}
\figurenum{A1}
\epsscale{1.}
\plotone{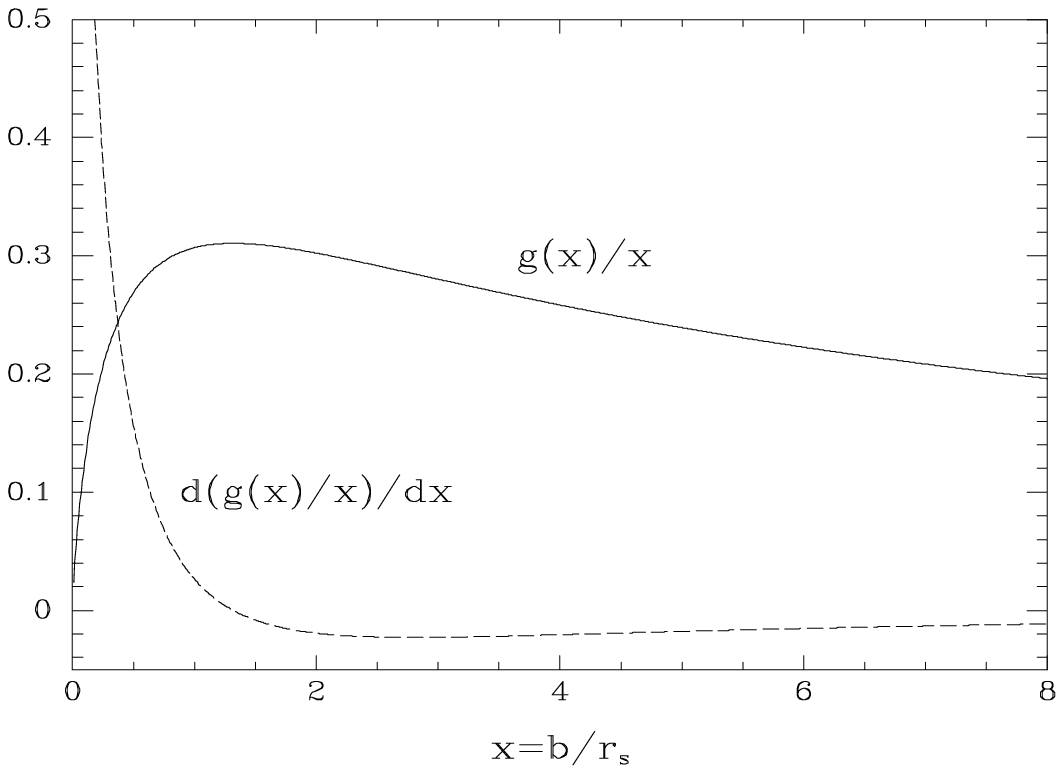}
\caption{The dimensionless function $g(x)/x$
which characterizes the bending angle of
a light ray passing at impact parameter
$x$ from a cluster with a Navarro, Frenk, \& White
profile (eqs. A5--A8). The dashed line is this function's
derivative.
}
\end{figure}

\end{document}